\documentclass[a4paper,10pt,english]{article}
\usepackage[margin=1in]{geometry}
\usepackage[utf8]{inputenc}
\usepackage{enumitem}
\usepackage[colorlinks]{hyperref}
\usepackage{graphicx}
\usepackage{color}
\usepackage{amsmath}
\usepackage{enumitem}
\usepackage{authblk}
\usepackage{gensymb}
\usepackage{setspace}
\usepackage{amssymb}
\usepackage{tablefootnote}
\usepackage[table,xcdraw]{xcolor}
\usepackage[
backend=biber,
style=numeric,
sorting=none
]{biblatex}
\title{Quantum Communication Countermeasures}
\author[1]{Michal Krelina\thanks{michal.krelina@cvut.cz}}
\affil[1]{Faculty of Nuclear Sciences and Physical Engineering, Czech Technical University in Prague, Brehova 7, Prague, Czech Republic}

\begin{document}
\maketitle

\begin{abstract}
Quantum communication, particularly quantum key distribution, is poised to play a pivotal role in our communication system in the near future. Consequently, it is imperative to not only assess the vulnerability of quantum communication to eavesdropping (one aspect of quantum hacking), but also to scrutinise the feasibility of executing a denial-of-service attack, specifically, stopping quantum communication from working. 
Focusing primarily on the free-space quantum channel, the investigation of possible denial-of-service attacks from a strategic perspective is performed. This encompasses the analysis of various scenarios, numerical modelling, risk estimation and attack classification.
The out-of-FOV (field of view) attack emerges as a particularly severe threat across nearly all scenarios. This is accompanied by proposed counter-countermeasures and recommendations.

\end{abstract}

%\begin{keyword}
\textbf{Keywords:} quantum communication, quantum countermeasures, QKD, laser, denial-of-service attack, quantum vulnerabilities 
%\end{keyword}

%{\color{red} Version 0.3 - basic proofreading, after review}

%---------------------------------------------------------------------------
%---------------------------------------------------------------------------

\section{Introduction}\label{sec:intro}

Quantum network and its protocols provide information-theoretic provably secure communication, including the exchange of encryption keys via quantum key distribution (QKD) \cite{BB84,Ekert1991,Tomamichel2017}. More advanced quantum information networks or quantum internet \cite{Wehner2018} can go beyond QKD and offer additional services such as quantum digital signatures \cite{Gottesman2001}, secure identification \cite{Damgrd2014}, secret sharing \cite{Han2008}, or quantum secure direct communication \cite{Zhang2017,Qi2019}.
Quantum communication has the potential to become a new security standard alongside post-quantum cryptography (PQC) to protect against attacks by future quantum computing and its quantum cryptoanalytical capabilities \cite{Shor1994,Grover1996,Bernstein2010}.

%{\color{purple}
%Recently, quantum communication might be felt as a stronger solution as more and more news about which PQC schemes have been broken, e.g. \cite{cryptoeprint2022}. 
%This holds even despite the fact that quantum communication requires building new infrastructures and generally needs to mature. The reward can then be additional non-security services such as distributed quantum computing \cite{Crpeau2002,Cuomo2020}, clock synchronisation \cite{Kmr2014}, or networked quantum sensors \cite{Proctor2018}.
%}

In comparison to PQC, quantum communication necessitates the establishment of new infrastructures and typically requires a period of maturation. The ensuing benefit, however, encompasses additional non-security services, including distributed quantum computing \cite{Crpeau2002,Cuomo2020}, clock synchronisation \cite{Kmr2014}, and networked quantum sensors \cite{Proctor2018}.
This is particularly evident in the case of integrated Quantum Information Networks (QIN), which are instrumental for long-distance quantum communication. These networks deploy various quantum assets such as quantum repeaters, quantum entanglement distribution systems, quantum sensors, and imaging systems in space around the Earth \cite{deForgesdeParny2023}.
These quantum assets manifest primarily as dedicated quantum satellites or integral components of space experimental laboratories. In the long run, we anticipate the seamless integration of quantum systems with classical counterparts, including laser or microwave communication systems, Earth surveillance systems, and so forth \cite{Krelina2023spa}.

One of the crucial advantages of classical communication over quantum is the ability to reroute traffic in the event of a line or node being severed, typically without major consequences. However, in the near to medium term, especially when quantum networks are not yet large-scale networks with numerous interconnected nodes, an attack on a quantum network asset in space could result in a significant denial-of-service (DoS) scenario. This could have far-reaching implications if critical communication infrastructure depends heavily on this link.

Quantum hacking \cite{Bugge2014,Makarov2016,Sajeed2016,Qian2018,Fei2018,Huang2018,Huang2018b,Huang2020,Chaiwongkhot2022} is a dynamic area of research focused on uncovering vulnerabilities in quantum communication that could potentially lead to eavesdropping on communications, which is often the primary objective.
Nevertheless, DoS attacks are, in principle, more straightforward, and it is vital to comprehend their nature and implications prior to quantum communication becoming an integral part of critical communication infrastructure. In such a scenario, we can draw parallels to electronic warfare and discuss quantum countermeasures \cite{Krelina2021,Krelina_2023,Krelina2023spa}.

Countermeasures in quantum communication, particularly in terms of dazzling or, albeit less accurately, jamming, remain a relatively unexplored area. There are primarily two works that delve into this topic in depth \cite{Gozzard2021, Simmons2023}. These works operate at the "tactical" level, focusing on specific scenarios, considering specific designs and offering analyses of QKD performance in terms of quantum bit error rate (QBER).

In this study, our aim is to approach this subject more strategically, with the objective of identifying potentially promising scenarios and vectors of attack. These will then be subjected to further detailed study and assessment. Within this context, we will examine the feasibility and types of attacks on quantum communication in free space, categorise types of DoS attacks, provide preliminary risk estimates, conduct initial numerical modelling, ascertain the prerequisites for such attacks, and propose potential mitigations.

%In this report, we investigate the possibility and types of attacks on quantum communication in free space using laser weapons. The goal is to classify types of DoS attacks using lasers, understand the requirements for such laser weapons and their availability, suggest countermeasures, and carefully define further tasks for more detailed studies.

This paper is divided as follows. First, Section~\ref{sec:qcomm} introduces quantum network design and security aspects. Section~\ref{sec:definitions} introduces the term "quantum communication countermeasure," describes the most vulnerable elements in a quantum network, provides basics on laser weapons, and classifies types of laser attacks on a quantum receiver. Then, Section~\ref{sec:laserAttack} provides boundaries of considered attacks on quantum communication and considered scenarios, and describes numerical modelling and risk estimation. Results are presented in Section~\ref{sec:results}, including their discussion. Section~\ref{sec:cc} then presents possible counter-countermeasures to protect quantum assets. Lastly, Section~\ref{sec:conclusions} concludes the paper. Appendix~\ref{sec:appx1} provides details on laser propagation.

%---------------------------------------------------------------------------
%---------------------------------------------------------------------------
%---------------------------------------------------------------------------
\section{Quantum communication security/resilience}\label{sec:qcomm}

In general, quantum information networks allow the exchange of quantum bits at large distances, ranging from laboratory settings to intercontinental spans. This enables a multitude of applications, including Quantum Key Distribution (QKD), distributed quantum computing, networked sensing, and other Quantum Information Network (QIN) services \cite{Wehner2018}.
The true service capability of a given QIN hinges largely on its ability to distribute quantum entanglement (quantum entangled quantum bits). If it can, numerous additional protocols and services mentioned above can be implemented.
If not, such a quantum network is usually considered for QKD only (no quantum entanglement is needed). 

A fundamental property of quantum communication lies in its inherent resistance to eavesdropping, provided certain conditions are met, and the entire system is properly implemented, rendering it information-theoretically secure \cite{Renner2005}. Consequently, the domain of security proofs for quantum communication protocols constitutes a dynamic field of research, focused on analysing the resilience of quantum protocols against various attacks, colloquially referred to as quantum hacking. Quantum hacking may target hardware imperfections (detectors, emitters) \cite{Bugge2014,Qian2018,Fei2018,Chaiwongkhot2022}, exploit vulnerabilities in the quantum link (e.g. many potential side channels for signals from coherent sources) \cite{Huang2018b,Huang2020}, or address imperfect implementations of quantum protocols themselves \cite{Sajeed2016,Huang2018}.
While most of these studies share the common objective of demonstrating the potential for eavesdropping on quantum communication (typically QKD, though many of these methods can be extended to other quantum communication protocols), they also propose mitigation for any identified vulnerabilities. 

In some cases,  these studies also discuss the possibility of (temporary or permanent) DoS attacks as a consequential outcome \cite{Bugge2014,Qian2018,Makarov2016}. Here, we are interested exactly in these DoS attacks. 
This paper specifically centres on these DoS attacks. In a severe conflict, interfering with and eavesdropping on an opponent's communication would provide a significant advantage. Yet, if such activities prove infeasible or overly challenging, a DoS attack, as part of electronic or information warfare, would be highly desirable and, indeed, anticipated \cite{van_niekerk_future_2009}.  

It's worth noting that many quantum side-channel attacks can be mitigated through improved hardware, such as single photon sources instead of weak coherent sources, adding additional optical elements (such as spectral and spatial filtering in the case of free-space communication), or new employing new protocols such as (measurement-)device-independent protocols whose proper implementation effectively eliminates most of the side channels. However, these mitigations are irrelevant in our context, where we consider intense illumination (blinding, dazzling), or temporary or permanent harm to optical or electronic components.

%---------------------------------------------------------------------------
%---------------------------------------------------------------------------
\subsection{Quantum network designs}

A comprehensive understanding of quantum network designs is crucial for a successful DoS attack. 
Quantum networks are implemented using either fibre optic links or free-space links. It is worth noting that a quantum network cannot function in isolation; classical channels for command and control are essential for its operation.

The fibre optic channel, commonly employed for terrestrial applications, is technologically more straightforward. However, the distance over which quantum information can be transmitted is constrained due to the exponential increase in transmission losses with distance, and amplification of the quantum signal is impossible due to the no-cloning theorem.
Presently, commercially available point-to-point connections can span up to 100 km \cite{Huttner2022} (although, through experimental utilisation of the so-called twin-field QKD protocols, distances of up to 1,000 km have been achieved \cite{Liu2023}). For greater distances, repeaters (relay points) become indispensable. In configurations involving three or more endpoints, or in more complex quantum network topologies, additional elements like optical switching or splitting may also be employed.

Repeaters can be categorised into trusted repeaters, which are relay nodes with access to the key and are assumed to be secure against intrusion and attacks by any unauthorised parties (hence, they are deemed ``trusted''), and quantum repeaters (quantum relay nodes) equipped with quantum memory and/or capable of quantum entanglement distribution. 

\begin{figure}[htb]
    \centering
    \includegraphics[width=\textwidth]{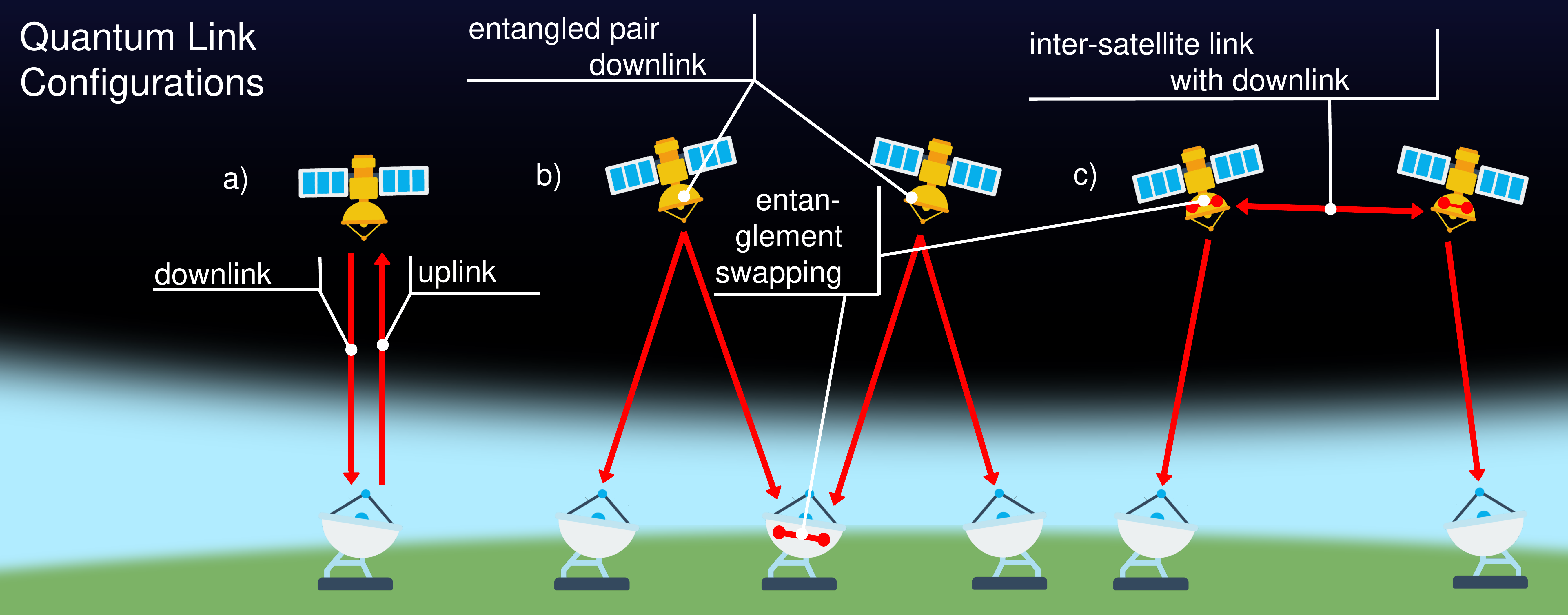}
    \caption{Various considered quantum link configurations. Configuration a) simple downlink and/or uplink considered for QKD. Configurations b) and c) are considered for quantum information networks with quantum entanglement. Here, b) is only with downlinks and entanglement swapping at the ground. c) is with downlink and inter-satellite link with entanglement swapping at satellites.}
    \label{fig:space-links}
\end{figure}

The free-space link comprises multiple optical ground stations (OGSs) and quantum satellites acting as relay points. Fig.~\ref{fig:space-links} illustrates the three most common designs for free-space quantum links:
\begin{itemize}
    \item[a)] Simple downlink and uplink.
    \item[b)] Entangled downlink (involving the transmission of two entangled photons towards two OGSs) with the potential for entanglement swapping at the OGS.
    \item[c)] Downlink with bidirectional inter-satellite links (with potential entanglement swapping occurring at the satellites).
\end{itemize}
OGSs can be seamlessly integrated into terrestrial quantum networks. Future concepts for leveraging quantum communication in space envision even more complex designs and topologies, including the utilisation of intermediate quantum satellites \cite{Gonzalez-Raya2023}.
In particular, space-based free-space quantum links enable communication over distances of thousands of kilometres \cite{Yin2017} owing to the general square-law relationship compared to transmission in fibre links. As will be detailed later, the space-based quantum link offers additional advantages for the downlink, as the primary losses in the atmosphere occur within the final approximately ten kilometres \cite{deForgesdeParny2023}.  
Conversely, ground-to-ground free-space links have been experimentally tested with a range of up to 144 km \cite{SchmittManderbach2007} due to atmospheric losses.

Experimental testing of QKD in space encompasses simple downlinks from Low Earth Orbit (LEO) \cite{Liao2017} or Geostationary Orbit (GEO) \cite{Wang2021}, uplinks to LEO \cite{Ren2017}, or entangled pair downlinks \cite{Yin2017,Yin2017b}. The majority of presently planned QKD projects, whether for demonstration or nearing commercialisation, focus on either a simple downlink or an entangled downlink at LEO \cite{Oi2017,Villar2020,Lewis2021,Eagle1}. However, there are also exceptions, such as \cite{Kerstel2018}, which explore the possibility of an uplink.

%In general, space and terrestrial quantum networks slightly differ in considered applications since the terrestrial quantum networks are expected to be faster (expected higher delivery key rate) than the space quantum network due .

%---------------------------------------------------------------------------
%---------------------------------------------------------------------------
%---------------------------------------------------------------------------
\section{Quantum communication countermeasures}\label{sec:definitions}

For the purpose of this report, the term ``quantum communication countermeasures'' (QCCM) refers to various actions leading to temporary or permanent denial of quantum communication service, or its significant slowing down. We identified that quantum communication countermeasures could be conducted in three vectors:
\begin{enumerate}
    \item Quantum link - free-space link
    \item Quantum link - fibre optic link
    \item Classical link - quantum networks command and control 
\end{enumerate}

The attack on classical links can be executed through a wide spectrum of cyber methods \cite{Haig2015}, and as such, will not be extensively covered here, with a few exceptions.
From this perspective, a satellite for quantum communication is a conventional satellite equipped with quantum-optical, optical, and radio frequency (RF) payloads. Therefore, all countermeasures designed against classical satellites can be applicable here. This encompasses kinetic anti-satellite (ASAT) weapons, cyber attacks, radio-frequency jamming, as well as high-energy laser, microwave, and particle beam attacks \cite{Bohek2022,Colby2016}.

The attack on fibre optic links poses a greater challenge, requiring prior knowledge of the location of the optic fibres or quantum nodes for a successful execution. In this context, a simple fibre interruption suffices. However, underwater optical fibres are of heightened interest, presenting a greater probability of sabotage in case of conflict \cite{Krelina2023spa}. In the early stages of quantum networks, such an attack can prove highly effective due to the network's low density, making rerouting very costly or even impossible.
As simulated in \cite{Cicconetti2022}, quantum network resources escalate exponentially, while fidelity experiences an exponential decrease with path length. 

The scenario becomes more intriguing in the case of the free-space channel, where, under specific conditions, multiple strategies can be employed. In the following, we consider the laser as the most viable tool for use against quantum channels or optical communication in general.

%---------------------------------------------------------------------------
%---------------------------------------------------------------------------
\subsection{Quantum network elements in space}

The fundamental components of quantum free-space communications are the receiver and the transmitter. 
Broadly put, a transmitter comprises a source of single photons or weak coherent pulses, optical elements (mirrors, filters, attenuators, beam splitters, etc.), and a telescope. Conversely, the receiver generally encompasses a telescope, optical elements, and a single photon detector (SPD). Typically, the integral parts of both the receiver and transmitter include components for the beacon laser (emitter and receiver) as part of the acquiring, pointing, and tracking (APT) system, which also incorporates an additional laser and a camera (such as a CCD). A typical representation of a receiver and transmitter can be found in Figure~1 in \cite{Liao2017}.

For the purposes of this discussion, we consider the use of a laser for the attack unless otherwise specified. 
When contemplating a laser attack, we identified the single photon detector as the most susceptible target for QCCM based on the literature review and simulations. 

The SPD is, in essence, an extremely sensitive device capable of detecting individual photons within a specific wavelength range. Two main types of SPDs are considered: semiconductor-based avalanche photodiodes (APDs) and superconducting single-photon detectors (SNSPDs). While SNSPDs generally exhibit superior performance, they are more expensive and necessitate cryocooling compared to APDs \cite{deForgesdeParny2023}. 
Thresholds for blinding or damaging APDs have been studied, for instance, in \cite{Bugge2014,Makarov2016,Lydersen2010,Erlong2005}, although comprehensive studies on laser damage to APDs have not been conducted yet. This differs from the situation with camera sensors \cite{Bartoli1977,Schwarz2017}. 
The values of laser power and their effects are presented in Tab.~\ref{tab:lws_effects}. It is essential to note that the presented laser power is applied directly to the SPD. In real Quantum Key Distribution (QKD) systems, several optical elements such as bandpass filters (BPS), beam splitters (BS), or polarising beam splitters (PBS) are placed prior to the SPDs, effectively reducing the collected laser power to individual SPDs.

\begin{table}[h!]
    \centering
    \begin{tabular}{c|l|l}
      Power / Power density & Effect & Ref. \\
      \hline
      $\gtrsim 10^{-15}$~W &  Too high noise for SPD & \cite{Erlong2005}
      \\
      $\gtrsim 10^{-11}-10^{-8}$~W &   Non-gated SPD APD blinding
      \tablefootnote{Depends on the bias voltage supply impedance.}  
      & \cite{Makarov2009,Lydersen2010} 
      \\
      $\sim 10^{-3}$~W &   APD thermal blinding & \cite{Lydersen2010}
      \\
      $\gtrsim 10^{-1}$~W/cm$^2$  &   CCD image transducer saturation threshold (used as part of APT) & \cite{ChinaASAT07}
      \\
      $\gtrsim 1.2$~W &   APD permanent blinding, lower sensitivity & \cite{Bugge2014} 
      \\
      $\gtrsim 2$~W &   APD structural damage, complete insensitivity & \cite{Bugge2014}
      \\
      $\gtrsim 4$~W &   Attenuators damage\tablefootnote{Optical attenuators are used for the preparation of weak coherent states. Their damage can lead to a decrease in attenuation, i.e. to higher mean photon numbers in the prepared states, allowing an eavesdropper to compromise the key.}  &
      \cite{Huang2020}
      \\
      $\gtrsim 3$~W &   Polarisation spatial filter degradation &
      \cite{Makarov2016}
      \\
       $\gtrsim 3\times10^{2}$~W/cm$^2$  &   Optical glass melting & \cite{ChinaASAT07}
      \\
       $\gtrsim 10^{3}$~W/cm$^2$  &   Melting initiation threshold for aluminium  & \cite{Nielsen_1994}
    \end{tabular}
    \caption{
    Summary of the approximate laser power or power density needed for various effects. It is important to note that the minimal time of target illumination by a laser to cause the discussed effect was not studied in all cases. The power per area needed is to be transformed to power based on the receiving telescope area. 
    %For comparison, the intensity of sunlight in the micro-wave regime is about $0.06$~W/cm$^2$.
    }
    \label{tab:lws_effects}
\end{table}

The free-space optical communication has some specific features.
Firstly, one must take into account the field of view (FOV). 
Telescope systems used at satellites and OGSs, when communicating, must be in a mutual line of sight (in the best case with the same optical axis), i.e. in their respective FOVs. 
This is a critical consideration for our analysis, as any strong attack leading to the destruction of an SPD must be directed into the target's FOV.  
Given that the typical FOV angle ranges from 3~$\mu$rad (e.g. \cite{Oi2017}) to around 500~$\mu$rad (e.g. \cite{arxiv.2208.10236}), there is extremely limited space for an ``in-FOV'' attack, as demonstrated in Table~\ref{tab:FOV}. It is worth noting that the FOV for the APT system is on the order of a few mrad or tens of mrad.

On the other hand, due to receiver system imperfections, an ``out-of-FOV'' attack can be considered
as more relevant, as first suggested in \cite{Gozzard2021}.

\begin{table}[htb!]
    \centering
    \begin{tabular}{l|l|l|l|l}
%             & \multicolumn{4}{c}{FOV diameter $d$ [m] at the distance $l$ [km]}  \\
        $\phi_{FOV}$ [$\mu$rad] & $l=10$~km & $l=500$~km & $l=1,000$~km & $l=35,000$~km  \\
        \hline
        1 & 0.01 & 0.5 & 1 & 35 \\
        10 & 0.1 & 5 & 10 & 350 \\
        100 & 1 & 50 & 100 & 3,500 \\
        1,000 & 10 & 500 & 1,000 & 35,000 \\ 
    \end{tabular}
    \caption{FOV diameters $d$ [m] at distance $l$ for selected FOV angle.}
    \label{tab:FOV}
\end{table}

Secondly, photon loss due to atmospheric absorption and scattering is dominant only in the lower approximately 10 km. In general, two wavelengths are considered: 810~nm and 1550~nm.
Communication at 810~nm has a beam divergence one half of that at 1550~nm. However, 810~nm is overwhelmed during the day. 1550~nm is considered to be possibly used during daylight \cite{deForgesdeParny2023}. 

Thirdly, OGSs are usually positioned at fixed locations, lacking flexibility and potentially being vulnerable to constant surveillance and probes. Conversely, the protective elements of OGSs can be continuously improved compared to component replacements at a satellite in Low Earth Orbit (LEO) and above.

\subsection{Laser weapon systems}

As mentioned earlier, in our analysis, we will consider an attack on free-space quantum communication using a laser weapon system (LWS).
LWS can be divided into three types according to the impact \cite{Weeden_2022}:
\begin{itemize}
    \item \textbf{Dazzling} is considered more as a countermeasure than a weapon because it has no permanent effect. Typically, dazzling targets optical sensors like a charge-coupled device (CCD) or a complementary metal-oxide semiconductor (CMOS). In our context, it will mean the increasing level of noise leading to the DoS. 
    
    \item \textbf{Blinding or satellite image sensor damage} (or associated electronics and optics) is caused by a laser with higher energy. The damage is permanent and can refer to either single-pixel damage or damage to the entire sensor. In our case, it means temporary or permanent damage to SPD. 

    \item \textbf{Damage to the satellite bus} requires a very high-energy laser where the damage is considered due to the thermal effects of the absorbed energy. This attack is focused on critical systems like the thermal regulation system, batteries, solar panels, or the attitude control system, resulting in the complete failure of the satellite. Typically, the delivered energy for melting should be around and higher than 10,000 J \cite{Nielsen2012-gv} (e.g., 2~kW impinging for 5 seconds).
\end{itemize}

It's important to note that the threshold between dazzling and blinding is almost impossible to predict, which introduces a risk of blinding during the action of dazzling countermeasures.

Counterspace or anti-satellite LWS encompasses not only lasers but also high-fidelity space situational awareness, high-power laser devices, precise beam tracking and control, and adaptive optics to counteract atmospheric turbulence (for ground-based lasers) \cite{Weeden_2020}.

In general, (space) laser weapons are a well-studied topic, as seen in references such as \cite{Altmann88, Titterton2015-qr}. However, detailed information and parameters of current and currently developed LWS are naturally classified. Available information is usually limited to the range of output power, and the aperture can be deduced from published photos. Some overview of the current status in the counterspace sense can be found in \cite{Weeden_2022,BLAstakK_2022,ChinaASAT07,Liu_Lin_Chen_2020}.

%---------------------------------------------------------------------------
%---------------------------------------------------------------------------
%---------------------------------------------------------------------------
\section{Attacks against quantum assets}\label{sec:laserAttack}

Within this section, we will set boundaries for our analysis and specify individual scenarios along with their objectives.

\begin{figure}[htb!]
    \centering
    \includegraphics[width=0.7\textwidth]{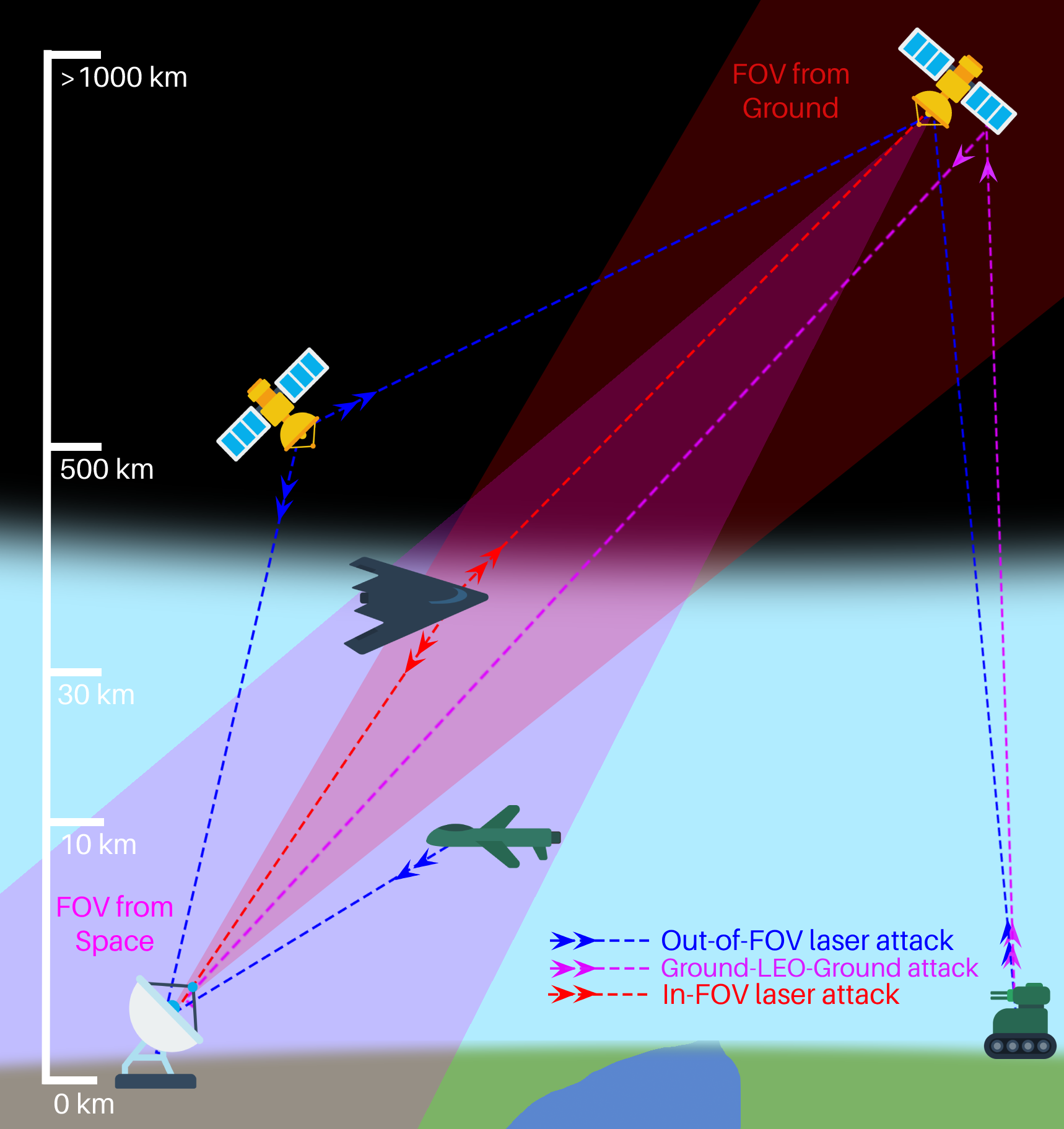}
    \caption{This picture illustrates the FOV of the satellite and OGS targets, encompassing possible in-FOV and out-of-FOV attacks, as well as the distinct Ground-LEO-Ground attack variation across various platforms.}
    \label{fig:space-attack}
\end{figure}

First and foremost, we categorise attacks into two types. The first is the in-FOV attack (direct attack), wherein the LWS is within the target's Field of View (FOV) and the entirety of the initial laser energy, minus losses, is directly channelled into the quantum receiver system. The in-FOV attack bears the potential for the quantum receiving system's destruction. Second, the out-of-FOV attack (side attack) takes into account receiver imperfections, specifically, the acquisition of a small fraction of laser light due to internal scattering. This form of attack, at its best, can achieve a dazzling effect. 
The possible ways of the in-FOV and out-of-FOV attacks are illustrated in Figure~\ref{fig:space-attack}.

Next, we consider only the delivered laser power in our analysis. This is a very simplified approach since critical is not only the Subsequently, our analysis exclusively considers the delivered laser power. This approach is notably simplified, as the critical factor is not only the transmitted power but also the transmitted energy, i.e., the duration the target was illuminated. Another crucial aspect is whether the employed LWS was a continuous wave (CW) or a pulsed laser. Furthermore, different destruction mechanisms come into play depending on whether nano- and picosecond or femtosecond laser pulses are employed \cite{ristau2014laser}. However, we exclude consideration of femtosecond lasers due to additional non-linear effects in atmospheric propagation.

The existing literature presently lacks the depth required for a more detailed analysis concerning quantum communication components. To illustrate, in \cite{Bugge2014}, it is reported that a 2~W CW laser led to the destruction of APDs when illuminated for 60~s. However, it is important to note that 60~s may not represent the minimal time required for permanent damage to occur in APDs.

Additionally, we refrain from specifying a particular Laser Weapon System (LWS) in terms of output power properties or the employed laser technology. Instead, we present a universal output and assessment of LWS initial power, approximating from 1~W to 1~MW. 
Unless explicitly stated, we assume that the LWS is configured to mitigate or suppress non-linear atmospheric effects such as blooming \cite{Gebhardt1976}. This implies a continuous-wave laser of up to 25~kW (where the blooming effect begins to exhibit significance \cite{Zohuri2016}) when operated from a ground-based platform.

For the sake of simplicity, we consider a zenith angle of $\phi=0^{\degree}$ (or an elevation angle of $90^{\degree}$) for the in-FOV attack. This implies that the LWS is always directed perpendicular to the horizon plane, representing the shortest path through the atmosphere and to the target (i.e., altitude, $h$, equates to the propagation path, $L$). Conversely, for out-of-FOV scenarios, we assume a zenith angle of $\phi=60^{\degree}$, simulating an attack from a greater distance, such as from a neighbouring country.
It is worth noting that a zenith angle of $\phi=60^{\degree}$ results in a suppression factor of approximately $\sim 0.37$ for a satellite at an altitude of $h=1,000$~km, and approximately $\sim 0.58$ at an altitude of $h=500$~km when compared to $\phi=0^{\degree}$.

For both types of attack, we consider an ideal Laser Weapon System (LWS) targeting system, wherein the axis of the laser beam emitted from the LWS is precisely aligned with the centre of the target's aperture, in conjunction with high-fidelity space situational awareness.
In the case of the out-of-FOV attack, the simulation was based on the in-scattering profile \cite{Gozzard2021}, with a slight modification incorporating a suppression parameter $\kappa_{outFOV}$ to account for increased uncertainty stemming from potential future enhancements in design, attack angles, and related factors. Further details can be found in Appendix~\ref{sec:appx1}.

As a result, this analysis serves as an illustrative approximation - a strategic perspective, offering a magnitude estimate of the requisite LWS power, and it yields insights into potentially significant scenarios.

An alternate method for executing a quantum DoS attack involves the generation and dispersion of scattering aerosols, such as soot aerosol \cite{Zhang2022}. This would necessitate dispersal above the OGS, potentially from an aircraft. However, as described later in more detail, we consider this scenario improbable due to anticipated anti-aircraft defence measures, resulting in only a limited number of plausible scenarios.

A potent variant of the out-of-FOV attack, not involving lasers, is the ``redout'' effect triggered by the detonation of a nuclear weapon \cite{Wilson2022}. In this context, we do not contemplate the use of nuclear detonations solely for the purpose of executing a quantum communication DoS.

Finally, our analysis is also simplified regarding the definition of quantum communication operability. Many studies on quantum communication focus on parameters like key rate (in the case of QKD) or universally on the QBER. The QBER typically has a threshold, commonly around $8-12\%$, beyond which, due to high noise levels, the communication is deemed insecure and terminated. Our approach is derived from \cite{Erlong2005}, establishing a minimal received power, $P_{recv} \gtrsim 10^{-15}$~W, where noise levels become prohibitively high for secure quantum communication. This minimal receiver power is established based on limits on the number of noise/background photons, at $3\cdot10^{-5}$~count pulse$^{-1}$, to maintain an error rate below $5\%$. It's important to note that elevated noise levels, which approach the QBER threshold, also result in a reduction in the quantum communication rate.

%---------------------------------------------------------------------------
%---------------------------------------------------------------------------
\subsection{Scenarios}\label{sec:scenarios}

Here, we specify various scenarios with different vectors of attack. Additionally, we assume that the positions of ground stations as well as satellites are known. 
In brief, quantum communication satellites can be tracked by space surveillance. The surveillance of space objects and debris is typically carried out by ground sensors, such as the Space Fence project \cite{Erwin2020}, or by the EU Space Surveillance and Tracking (SST) support framework \cite{EUSST}.
Determining the location of ground stations poses a greater challenge. They can be localised in the case of using an RF classical link for command and control. Otherwise, we must rely on intelligence work.

We consider a GEO satellite at an altitude of $h=35,800$~km, LEO satellites at altitudes of $h=500$ and $1,000$ km. For the LWS in the air domain, we contemplate drones, planes, or stratospheric vehicles at typical altitudes of $h=5$, $10$, and $30$ km. Regarding the ground component, both LWS (mobile - ships, trucks, and fixed) and OGSs correspond to an altitude of $h=0$.

It's worth noting that for the in-FOV attack originating from the ground, we also consider the use of adaptive optics (AO) to minimise atmospheric losses. On the other hand, out-of-FOV attacks, especially with high power, may not require AO and possibly only a coarse targeting system.
  
In our calculations, we assume good (cloudless) weather conditions. Nonetheless, it's important to recognise that adverse weather will typically have a more detrimental effect on quantum communication compared to a high-power laser beam.

We consider the following scenarios:
\begin{itemize}
    \item \textbf{Ground-LEO-Ground} scenario involves a strong illumination of an LEO satellite, causing a~significant portion of the laser to be reflected towards the ground-based receiver. In this scenario, the satellite behaves like a very bright but moving star, as first described in \cite{Gozzard2021}. The LWS can be a mobile (truck or ship) or fixed LWS facility, with fewer limitations on available power.

    \item \textbf{Ground-LEO} scenario involves the in-FOV or out-of-FOV attack, where a ground-based LWS (facility, ship, or truck) potentially employs a high-energy laser for a minimal time in the target's FOV to achieve a direct hit. The success of the out-of-FOV attack is very probable, even with low LWS output power.

    \item \textbf{Ground-GEO} scenario is similar to the Ground-LEO scenario. On one hand, the situation for the attacker is easier due to the stable and known position of the GEO satellite. On the other hand, the required laser power is higher due to the intensity decrease as $1/L^2$ (compare the suppression $\sim 8\times 10^{-10}$ for GEO with $\lesssim 10^{-6}$ for LEO), where $L$ is the distance. Therefore, this scenario is considered valid, especially for the out-of-FOV attack.
 
    \item \textbf{Air-Ground} scenario considers an LWS placed on a plane, drone, or stratospheric vehicle. Due to the relatively small distance between the LWS and the ground target, i.e., a small FOV cone, conducting the in-FOV attack is very challenging. The out-of-FOV attack is considered quite accessible.
    
    \item \textbf{Air-LEO} scenario is similar to the previous one except for targeting the LEO satellites. The in-FOV attack is again very challenging. The potential advantage could be reduced losses due to the propagation in the atmosphere and simplified LWS without adaptive optics from approximately 5~km or higher.
  
    \item \textbf{LEO-Ground} considers an LWS placed at the LEO satellites targeting OGS. Here, the in-FOV attack is disadvantaged with relatively low available laser power and high speed (more than $7,700$~m/s at $h=500$~km), allowing for an extremely short time window inside the target's FOV considering that the satellite's trajectory leads through FOV - which will be rather a very rare case. On the other hand, the out-of-FOV attack on the target is more accessible.

    \item \textbf{LEO-LEO} scenario could be relevant for more advanced network designs where the inter-satellite quantum link with entanglement distribution and swapping is considered. However, this scenario could be very challenging for the attacker due to the relatively fast mutual motion and the need for advanced orbital manoeuvrability of the satellite with the LWS payload. Nevertheless, the out-of-FOV attack could be considered more feasible.

    \item \textbf{LEO-GEO} scenario, similar to LEO-LEO, can be expected to be technically very challenging. Due to the payload limitations, mostly the out-of-FOV attack (where low laser power is sufficient) can be expected.

    \item \textbf{GEO-Ground} scenario might have the advantage of a vast area within reach for (possibly permanent) laser illumination, making it particularly relevant for the out-of-FOV attack. The main limitation will be the available laser power.
    
\end{itemize}

Other scenarios, such as Air-GEO, and Ground-MEO (medium Earth orbit, between $2,000$ and $35,800$~km), can be generalised and interpolated from the scenarios considered above.

It's worth noting that a more precise simulation should also take into account the scattering of laser photons from the atmosphere, where a limited number of photons can be scattered inside the FOV. However, such a scenario was not investigated in this study.

%---------------------------------------------------------------------------
%---------------------------------------------------------------------------
\subsection{Numerical modelling}\label{sec:nummodelling}

In our simulation, several parameters had to be set. Firstly, the size of the apertures (of parabolic mirrors or lenses) were fixed. The specific sizes of apertures are detailed in Tab.~\ref{tab:apertures}, and they were determined as averages based on a survey of the relevant literature.

\begin{table}[!htb]
    \centering
    \begin{tabular}{l|c}
      System & Diameter [m] \\
      \hline
       Ground LWS  & 1.0 \\
       Airborne LWS  & 0.2 \\
       Space LWS  & 0.2 \\
       Ground quantum communication system  & 0.6 \\
       LEO quantum communication system  & 0.2 \\
       GEO quantum communication system  & 0.2 \\
    \end{tabular}
    \caption{Fixed averaged sizes of apertures of various considered systems.}
    \label{tab:apertures}
\end{table}

Next, we consider that the LWS will operate at wavelengths either identical to or very close to those used in quantum communication, specifically 810 and 1550~nm, in order to bypass band filters.

We take into account the parameters of the LWS to ensure avoidance or suppression of non-linear atmospheric effects such as thermal blooming. This typically involves the implementation of adaptive optics, active beam control, managing pulse duration and repetition rate, or employing a CW laser restricted to a certain power density threshold (which depends on wavelength, altitude, and atmospheric conditions) \cite{Zohuri2016}.

The characteristics of free-space and atmospheric propagation are detailed in Appendix~\ref{sec:appx1}, which includes modelling for adaptive optics.

It is crucial to note that our numerical calculations assume ``good'' weather conditions without clouds. In the case of cloudy weather, both the attacks and quantum communication itself are weakened.

For this simulation, we have developed our own simulation framework, \textsc{Aether} (currently at version 0.2), written in Python. \textsc{Aether} is a free-space propagation simulator comprising individual modules that represent distinct optical elements and environments. It not only simulates the free-space propagation of laser but is also prepared for the simulation of free-space quantum communication, such as QKD, in subsequent projects.

%---------------------------------------------------------------------------
\subsection{Risk estimation}\label{sec:RiskEst}

The risk is assessed through a combination of the probability of a particular attack and its potential impact, as outlined in Table~\ref{tab:risk}. These impacts are categorised as follows: Negligible to no effect, Marginal to dazzling, Critical to blinding, and Catastrophic to satellite bus damage.

Based on this assessment, the risk can be classified into four categories, each with corresponding recommendations:
\begin{itemize}
    \item \textbf{Low risk} - Continue with existing controls, but closely monitor for any changes.

    \item \textbf{Medium risk} - Requires focused attention to mitigate the risk, along with regular or continuous monitoring. It is advisable to implement design-originated mitigations.

    \item \textbf{Serious risk} - Demands immediate action to reduce the risk to an acceptable level. It is imperative to implement a comprehensive set of countermeasures, including considering alternative routes for quantum communication.

    \item \textbf{High risk} - The risk is deemed excessively high and is not acceptable. Rigorous mitigation measures must be put in place. There is a significant risk of unfeasible quantum communication in at least the medium to long term.
\end{itemize}

\begin{table}[htb!]
    \centering
\begin{tabular}{|l|l|l|l|l|}
\hline
\textbf{Likelihood/Impact} & \textbf{Negligible}            & \textbf{Marginal}               & \textbf{Critical}               & \textbf{Catastrophic}           \\ \hline
\textbf{Improbable}        & \cellcolor[HTML]{32CB00}Low    & \cellcolor[HTML]{32CB00}Low     & \cellcolor[HTML]{F8FF00}Medium  & \cellcolor[HTML]{F8FF00}Medium  \\ \hline
\textbf{Remote}            & \cellcolor[HTML]{32CB00}Low    & \cellcolor[HTML]{F8FF00}Medium  & \cellcolor[HTML]{F56B00}Serious & \cellcolor[HTML]{F56B00}Serious \\ \hline
\textbf{Probable}          & \cellcolor[HTML]{F8FF00}Medium & \cellcolor[HTML]{F56B00}Serious & \cellcolor[HTML]{F56B00}Serious & \cellcolor[HTML]{FF0000}High    \\ \hline
\textbf{Frequent}          & \cellcolor[HTML]{F8FF00}Medium & \cellcolor[HTML]{F56B00}Serious & \cellcolor[HTML]{FF0000}High    & \cellcolor[HTML]{FF0000}High    \\ \hline  
\end{tabular}
    \caption{4x4 risk assessment matrix.}
    \label{tab:risk}
\end{table}

The likelihood of an attack was assessed differently for each scenario, based on the expected opportunities for a laser attack.
For instance, in the case of a Ground-to-LEO attack, the in-FOV attack was considered improbable due to the very low chance of the ground-based LWS entering the target's FOV cone. However, the out-of-FOV attack was deemed probable, considering certain range and weather limitations.
Another example is an Air-to-Ground/LEO attack involving various LWS platforms such as drones, planes, and stratospheric vehicles. While the in-FOV attack might be slightly more probable than a Ground-to-LEO attack, it is still highly improbable to achieve target acquisition within the small FOV (approximately $\phi_{FOV}=10$ $\mu$rad, see Table~\ref{tab:FOV}). The out-of-FOV attack is considered more probable than the Ground-to-LEO attack due to the significantly higher mobility of the LWS.

In particular, the following values were considered:
\begin{itemize}
    \item \textbf{Ground LWS:} Available power ranging from 1~kW to 1~MW.
    \item \textbf{Drone LWS:} Speed approximately 150~km/h; altitude about 5~km; laser power range 0.1-2~kW.
    \item \textbf{Plane (e.g. C-17 Globemaster) LWS:} Speed approximately 830~km/h; altitude about 10~km; laser power range 1-100~kW.
    \item \textbf{Stratospheric vehicle LWS:} Speed approximately 50~km/h; altitude about 30~km; laser power range 0.1-2~kW.
    \item \textbf{Satellite LWS:} Speed approximately $28,000$~km/h; altitude about 500~km; laser power range 0.1-2~kW.
\end{itemize}

%---------------------------------------------------------------------------
%---------------------------------------------------------------------------
\section{Results and discussion}\label{sec:results}

We conducted numerical modelling as described in Section \ref{sec:nummodelling} for each scenario, considering both in-FOV and out-of-FOV attacks. The results and selected numerical outputs are detailed and presented in the following subsection. Subsequently, we assessed the risk as outlined in Section \ref{sec:RiskEst}. 
 
Various potential mitigations are discussed in Section \ref{sec:cc}. However, it's important to note that many of these mitigations can be countered simply by employing an LWS with increased power, allowing the attacker to achieve a similar level of impact.

%---------------------------------------------------------------------------
\subsection{Scenario discussion}\label{sec:scenariores}

\begin{figure}[!htb]
    \centering
    \includegraphics{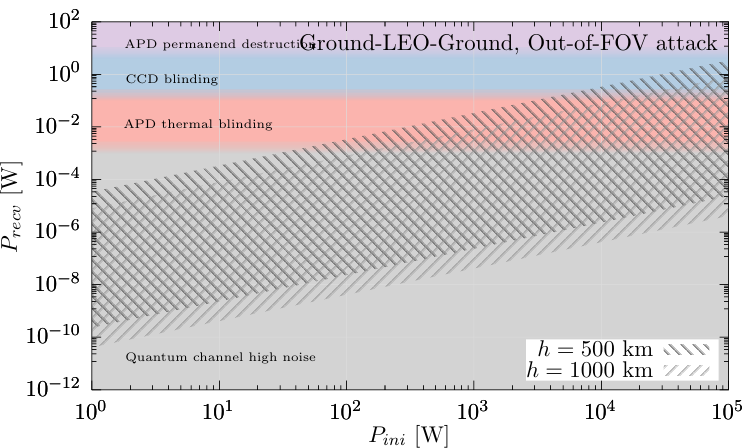}
    \caption{Initial LWS power, $P_{ini}$, versus received power at the target's quantum receiver, $P_{recv}$. The elevation angle between the LWS and the satellite is set at $30^\circ$, with zero zenith angle between the satellite and OGS. The hatched areas consider two extreme limits. The upper limit corresponds to a relevant satellite surface of 4 m$^2$ (such as for Micius) with an albedo of 1, while the lower limit assumes 0.01 m$^2$ (such as for a 1U CubeSat) with an albedo of 0.01, as described in Appendix~\ref{sec:appx1}.
    The background colours represent various effects - possibly no effect (white), too much noise for quantum communication (grey), APD thermal blinding (red), CCD blinding (blue), APD permanent destruction (purple), optics destruction (orange), and general melting (yellow).}
    \label{fig:Ground-LEO-Ground-out}
\end{figure}

\textbf{Ground-LEO-Ground scenario.} 
In this scenario, only the out-of-FOV attack is considered, where a ground-based LWS illuminates the transmitting QKD satellite, causing a fraction of the light to scatter towards the receiving OGS \cite{Gozzard2021}. The main advantage of this attack lies in the availability of high-energy lasers, ranging from tens to thousands of kW or even MW, for the ground LWS. Additionally, mitigating this attack is more challenging since the scattered light falls within the receiver's FOV. 
It was estimated in \cite{Gozzard2021} that an LWS with an output power of about 1 kW is sufficient to achieve the mission, i.e., quantum DoS, assuming a QKD satellite the size of Micius and a distance of about $1,000$ km between the LWS and OGS, which is very favourable for the attacker. Here, we show that even much lower laser power can be sufficient, as is illustrated in Fig.~\ref{fig:Ground-LEO-Ground-out}.

Given that the downlink form of space-ground communication is expected to be more common, it is reasonable to assume that this will be one of the most prevalent types of attacks.

The out-of-FOV attack can be mitigated, particularly through the proper design of the QKD satellite, which will have a small surface area (some designs operate with a size of 6U\footnote{The term ``6U'' refers to a design based on the size of six CubeSats, i.e., miniaturised satellites with a form factor of 10 cm cubes.}), along with a maximally reduced albedo.

It's important to note that the scattering model in \cite{Gozzard2021} is quite simplistic, and real-world scattering will vary for each satellite design. It will depend not only on the surface (CubeSat-based satellites are expected to have a smaller surface area) but also on the reflectivity of the material on the satellite's surface.

\textbf{Ground-LEO Scenario.} This scenario involves a direct attack on a quantum satellite in Low Earth Orbit (LEO) equipped with a quantum receiver. The out-of-FOV attack was described and modelled in \cite{Gozzard2021}. The concept here is to illuminate the target (LEO satellite) and utilise highly off-axis photons received by the telescope, which then get internally scattered and coupled into the QKD receiver.

\begin{figure}[!htb]
    \centering
    \includegraphics{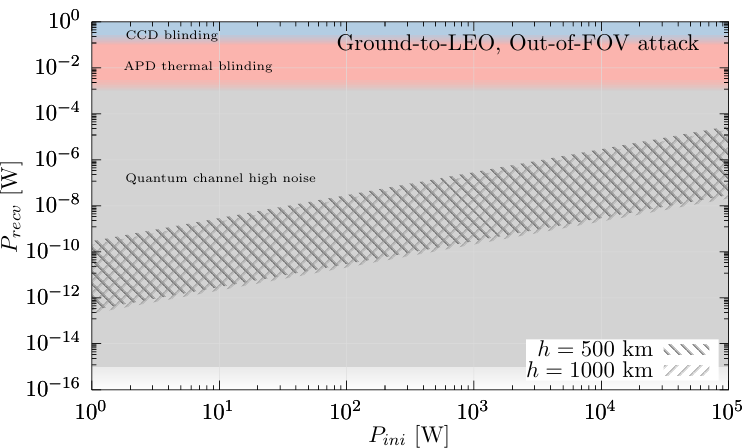}
    \caption{LWS initial power, $P_{ini}$, versus target received power impinging on the target's quantum receiver, $P_{recv}$, in the configuration of the out-of-FOV attack (with an elevation of $30^\circ$) for two distances between the LWS and the target. The hatched areas consider the range of $\kappa_{outFOV}$ over three orders of magnitude, as described in Appendix~\ref{sec:appx1}. The meaning of the background colours is the same as in Fig.~\ref{fig:Ground-LEO-Ground-out}.}
    \label{fig:Ground-to-LEO-out}
\end{figure}

Note that in this scenario, the out-of-FOV attack is considered less effective than the Ground-LEO-Ground scenario, primarily because the suppression parameter $\kappa_{outFOV}$ is less favourable for the attacker compared to the approximately additional $1/L^2$ factor corresponding to the loss during space-to-ground transmission after the reflection from the satellite. The potential performance of the out-of-FOV attack as a function of LWS initial power is depicted in Fig.~\ref{fig:Ground-to-LEO-out}. It is evident that even a 1~W laser, under good weather conditions and precise targeting, can lead to a DoS scenario.

The risk could be mitigated significantly, for instance, by improving the design of the receiving telescope, which serves as the main entry point for off-axis illumination.

As for the in-FOV attack from the ground, it is extremely challenging. With a realistic FOV angle of $\phi_{FOV}=10-500$ $\mu$rad, the ground LWS terminal would need to be positioned, for example, within approximately 5-250~meters of the transmitting OGS. This is an extremely challenging and highly improbable scenario. In the event of being within the FOV cone, the impact could be up to catastrophic, where 10~kW with Adaptive Optics (AO) or 100~kW without AO would be sufficient to permanently destroy the onboard SPD.

\textbf{Ground-GEO Scenario.} This scenario is, in principle, similar to the one described above, with the difference being the increased distance of approximately $40,000$ km (the LWS does not necessarily need to be placed on the equator). As a consequence, this scenario generally demands more energy, roughly by a factor of $1,000$, for the LWS to achieve similar results as in an attack on LEO satellites. In principle, the out-of-FOV attack can be continuous, and in cases of insufficient design, the risk could be classified as \textit{High}.

In theory, the satellite's FOV cone could extend up to several kilometres in radius on Earth. Nevertheless, we still consider it highly improbable for ground-based LWS (e.g., on a truck or boat) to reach it. Furthermore, achieving permanent destruction from the in-FOV attack would require several MW, which is unlikely due to the nonlinear effects in the atmosphere that significantly diminish the laser beam. Additionally, a MW laser would necessitate a large installation or facility.

\textbf{Air-Ground Scenario} involves placing the LWS on an airborne platform such as a drone (at an approximate altitude of $h \sim 5$~km), a plane ($h \sim 10$~km), or a stratospheric vehicle like a balloon ($h \sim 30$~km). While an LWS in the air could potentially enter the OGS's FOV cone, its feasibility is highly mission-dependent. For instance, it would be unlikely to execute such a mission within an opponent's territory where the anti-aircraft defence would be presented. This would be the most common scenario, which is why we lean more towards the \textit{Improbable} likelihood. On the other hand, it could be more feasible, for instance, in the case of no anti-aircraft defence, an OGS placed on a ship in international waters or directly on a battlefield.

\begin{figure}[!htb]
    \centering
    \includegraphics{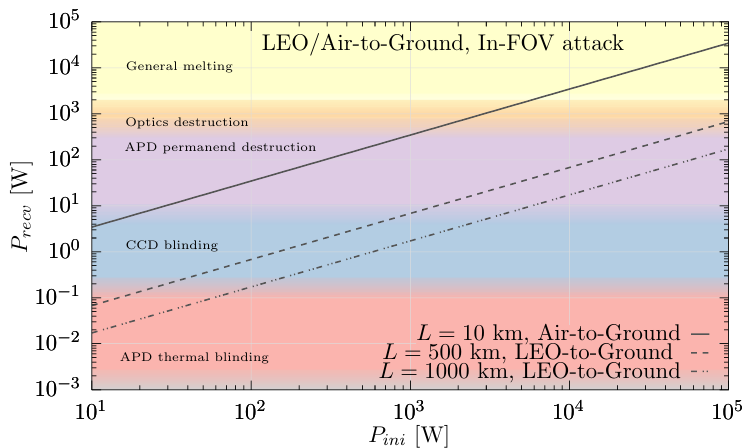}
    \caption{LWS initial power, $P_{ini}$, versus target received power, $P_{recv}$, for 810~nm at altitudes $L=10$ km (solid lines), $500$ km (dashed lines), and $1,000$ km (dot-and-dashed lines). The meaning of the background is the same as in Fig.~\ref{fig:Ground-LEO-Ground-out}.}
    \label{fig:AirLEO-to-Ground}
\end{figure}

In the potential case of the in-FOV attack, the effect of LWS power is visualised in Fig.~\ref{fig:AirLEO-to-Ground} (solid lines). Here, the required power for permanently damaging the SPD is about tens of watts, which is very accessible for all the mentioned airborne platforms. One could envision a special undercover mission with a drone flying into the target's FOV and damaging the OGS's quantum receiver.

The out-of-FOV likelihood was estimated as \textit{Remote} since even with only a $30^\circ$ elevation angle between the OGS and the stratospheric device, a horizontal distance of about 50~km from the OGS would still be required. We consider this as a remote probability due to the reasons described above - however, it is more probable than the in-FOV attack.

\textbf{Air-LEO Scenario} is similar to the previous scenario, with the target being the LEO QKD satellite. This effectively means a longer distance, and from an altitude of approximately $h \sim 10$ km, atmospheric effects can be neglected.

In this case, the longer distance results in larger geometric beam spreading, which could potentially be an advantage for the attacker, allowing for a global attack. For example, considering the out-of-FOV attack from an elevation angle of $30^\circ$ with a satellite at $h=500$~km implies a possible distance of about 850 km from the zero zenith angle, which is accessible for most potential missions.
 
The improbability of the in-FOV attack has the same reasoning as for the Air-Ground scenario.

\textbf{LEO-Ground Scenario.} This scenario is similar to the Ground-LEO scenario, but the LWS is placed in LEO. The main differences lie in the smaller beam widening of the laser (due to propagation in the atmosphere only in the last approximately 10 km) and the limited available laser power at the satellite. The fact that there's smaller diffraction can roughly compensate for the required energy for the out-of-FOV attack, which is available in space. For instance, in the Ground-to-LEO configuration, approximately five times higher initial power is needed to deliver the same laser power as in the LEO-to-Ground configuration, both considering $h=1,000$~km, different receiving aperture, etc. The out-of-FOV numerical results are depicted in Fig.~\ref{fig:Ground-to-LEO-out} (dashed and dot-and-dashed lines). Here, even 1~W initial power can be sufficient to cause dazzling on the Ground receiver.

\begin{figure}[!htb]
    \centering
    \includegraphics{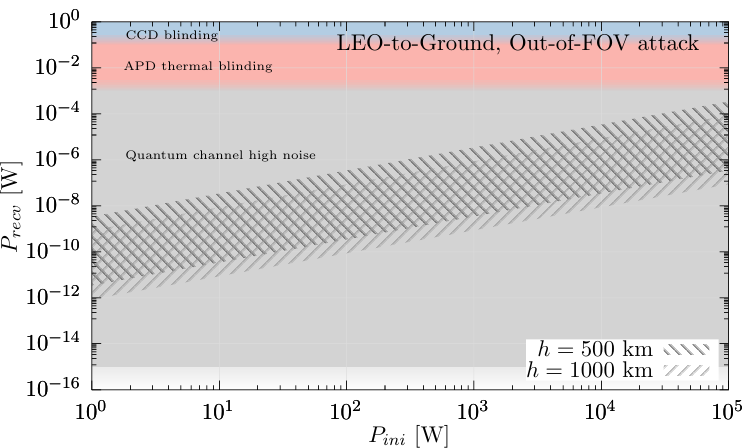}
    \caption{Same as Fig.~\ref{fig:Ground-to-LEO-out} except in the LEO-to-Ground scenario.}
    \label{fig:LEO-to-Ground-out}
\end{figure}

The in-FOV attack now is not limited by the potential anti-aircraft defence, and the LWS satellites can potentially cover the whole Earth. However, such LWS satellites would need to be numerous since active manoeuvring in LEO will be very fuel-consuming and unfeasible in a longer timescale. As demonstrated in Fig.~\ref{fig:AirLEO-to-Ground}, a space-based LWS at $h=500$~km would need approximately 2-3~kW laser power for SPD destruction, which is the upper limit of laser power considered for satellite-based lasers. Note that in the opposite case (Ground-LEO), for the same effect at the same distance, a power of more than 15~kW without AO or more than 3~kW with AO would be required.

\textbf{LEO-LEO scenario}.
This is technically a more complex scenario relevant for cases when the quantum receivers are present on satellites (uplink or inter-satellite links). Due to the absence of atmosphere, laser propagation and its effects are more straightforward. However, the in-FOV attack would have high requirements for manoeuvrability and fuel consumption.

The out-of-FOV attack, as in all other scenarios, is more accessible but also more challenging.

\textbf{LEO-GEO scenario} is another technically very challenging scenario, nevertheless considerable for the out-of-FOV attack where the missing propagation through the atmosphere (compared to Ground-GEO) is a significant advantage of a factor of about 4 that can even compensate for lower available laser power at LEO, i.e. from the ground you need a 4-times higher energy to case same effect at the same distance. Moreover, due to the fast orbiting (LEO satellites circle the Earth several times per day depending on the altitude), we consider the possible out-of-FOV attack \textit{remote}.

\begin{figure}[!htb]
    \centering
    \includegraphics{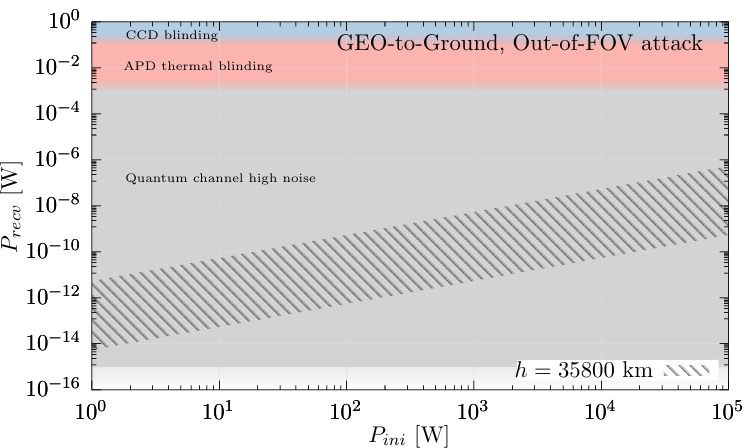}
    \caption{Same as Fig.~\ref{fig:Ground-to-LEO-out}, except for the GEO-to-Ground scenario.}
    \label{fig:GEO-to-Ground-out}
\end{figure}

\textbf{GEO-Ground scenario}.
Assuming limited laser power at GEO, we only consider the out-of-FOV attack, as shown in Fig.~\ref{fig:GEO-to-Ground-out}. The main advantage here is the potential to cover a vast portion of the Earth's surface. This implies that, in principle, we can execute permanent attacks.

The task in this scenario can be reformulated as follows: Let's assume a fixed initial laser power. Then, we can ask about the extent of the area that can be covered by received power $\ge 10^{-15}$~W to cause a DoS by varying the angle. Taking into account a simple geometry (i.e., zero zenith angle) and initial powers $W_{ini}=10$ and $100$~W, we can estimate a dazzled area with a radius of $r=1.1$ and $3.4$~km on the Earth's surface.

%---------------------------------------------------------------------------
\subsection{Risk estimation}\label{sec:riskres}

The primary results are presented in Table~\ref{tab:riskResults}, along with the corresponding risk classification as outlined in the risk assessment table (Tab.~\ref{tab:risk}) described in Section~\ref{sec:RiskEst}.

Our risk assessment did not identify any scenario where the risk could be classified as high, indicating that quantum communication should not be pursued at all. Conversely, the out-of-FOV type of attack is considered as \textit{Serious} in many of the studied scenarios. However, as discussed later, this risk can be mitigated, especially through proper and advanced design, which can lower the risk to \textit{Medium} or even to \textit{Low}.

\begin{table}[htb!]
    \centering
\begin{tabular}{|l|lll|lll|}
\hline
                  & \multicolumn{3}{c|}{\textbf{In-FOV attack}}                                                                              & \multicolumn{3}{c|}{\textbf{Out-of-FOV attack}}                                                                             \\ \hline
\textbf{Scenario}   & \multicolumn{1}{l|}{\textbf{Likelihood}} & \multicolumn{1}{l|}{\textbf{Impact}} & \textbf{Risk}                   & \multicolumn{1}{l|}{\textbf{Likelihood}} & \multicolumn{1}{l|}{\textbf{Impact}} & \textbf{Risk}                   \\ \hline
Ground-LEO-Ground & \multicolumn{1}{l|}{No}                  & \multicolumn{1}{l|}{No}              & None                            & \multicolumn{1}{l|}{Frequent}            & \multicolumn{1}{l|}{Marginal}        & \cellcolor[HTML]{F56B00}Serious \\ \hline
Ground-LEO        & \multicolumn{1}{l|}{Improbable}              & \multicolumn{1}{l|}{Catastrophic}    & \cellcolor[HTML]{F8FF00}Medium & \multicolumn{1}{l|}{Probable}            & \multicolumn{1}{l|}{Marginal}        & \cellcolor[HTML]{F56B00}Serious \\ \hline
Ground-GEO        & \multicolumn{1}{l|}{Improbable}              & \multicolumn{1}{l|}{Critical}        & \cellcolor[HTML]{F8FF00}Medium & \multicolumn{1}{l|}{Frequent}            & \multicolumn{1}{l|}{Marginal}        & \cellcolor[HTML]{F56B00}Serious \\ \hline
Air-Ground        & \multicolumn{1}{l|}{Improbable}             & \multicolumn{1}{l|}{Critical}        & \cellcolor[HTML]{F8FF00}Medium  & \multicolumn{1}{l|}{Remote}            & \multicolumn{1}{l|}{Marginal}        & \cellcolor[HTML]{F8FF00}Medium \\ \hline
Air-LEO           & \multicolumn{1}{l|}{Improbable}             & \multicolumn{1}{l|}{Critical}        & \cellcolor[HTML]{F8FF00}Medium  & \multicolumn{1}{l|}{Frequent}            & \multicolumn{1}{l|}{Marginal}        & \cellcolor[HTML]{F56B00}Serious \\ \hline
LEO-Ground        & \multicolumn{1}{l|}{Improbable}              & \multicolumn{1}{l|}{Marginal}        & \cellcolor[HTML]{32CB00}Low & \multicolumn{1}{l|}{Probable}            & \multicolumn{1}{l|}{Marginal}        & \cellcolor[HTML]{F56B00}Serious \\ \hline
LEO-LEO           & \multicolumn{1}{l|}{Improbable}             & \multicolumn{1}{l|}{Critical}        & \cellcolor[HTML]{F8FF00}Medium  & \multicolumn{1}{l|}{Remote}            & \multicolumn{1}{l|}{Marginal}        & \cellcolor[HTML]{F8FF00}Medium \\ \hline
LEO-GEO           & \multicolumn{1}{l|}{Improbable}             & \multicolumn{1}{l|}{Marginal}        & \cellcolor[HTML]{32CB00}Low     & \multicolumn{1}{l|}{Remote}            & \multicolumn{1}{l|}{Marginal}        & \cellcolor[HTML]{F8FF00}Medium \\ \hline
GEO-Ground        & \multicolumn{1}{l|}{Improbable}          & \multicolumn{1}{l|}{Marginal}      & \cellcolor[HTML]{32CB00}Low     & \multicolumn{1}{l|}{Frequent}            & \multicolumn{1}{l|}{Marginal}        & \cellcolor[HTML]{F56B00}Serious \\ \hline
\end{tabular}
    \caption{Risk assessment results for each scenario for in-FOV and out-of-FOV attacks.}
    \label{tab:riskResults}
\end{table}

%---------------------------------------------------------------------------
%---------------------------------------------------------------------------
%---------------------------------------------------------------------------
\section{Counter-countermeasures}\label{sec:cc}

The key parameter for free-space (quantum) communication is the signal-to-noise ratio (SNR). As quantum information is encoded in individual photons, the primary approach to enhancing SNR lies in noise suppression. To this end, incoming photons must traverse an array of optical devices before reaching a single-photon detector, which selectively filters out noisy photons. Such noise may originate naturally (e.g., from the sun, moon, or stars) or from man-made sources (such as city lights or reflections from satellites) \cite{Erlong2005}. 
In our context, these photons can also be regarded as elements of countermeasures, specifically, as the LWS photons that ultimately reach the detector.

Notably, the following improvements and risk mitigation elements are typically considered:

\begin{itemize}

    \item \textbf{Telescope Design.} The primary threat, as outlined earlier, is the out-of-FOV attack—receiving high off-axis photons. Therefore, whether placed on the ground, in the air, or in space, telescopes should be designed to limit these off-axis photons. This entails minimising scattering surfaces near the optics and augmenting internal baffling within the telescope \cite{Gozzard2021}. In this context, the higher energy of the LWS may be required to overcome this protective measure.

    \item \textbf{Satellite Architecture.} As described in the Ground-LEO-Ground scenario, the reflection of the laser from the transmitting satellite to the receiver poses a significant threat. Hence, the satellite bus should be engineered to minimise light reflection and scattering, which may involve employing suitable materials, such as anti-reflective coatings. However, this objective is partially at odds with the need for a highly reflective surface to address issues related to solar heating. In this context, the higher energy of the LWS may be required to overcome this protective measure.

    \item \textbf{Time-Gate Filter.} The SPD is in the ``on'' mode only when signal photons are anticipated. The efficiency of this process hinges on the short duration during which the gate is open, which, in turn, relies on the precision of synchronisation with the transmitting party—an aspect falling under the purview of the APT system. This mitigation is less effective against a CW laser, whereas a dazzling pulsed laser would need to align with the narrow time gate.

    \item \textbf{Wavelength Filter.} Much of the man-made background spans a broad wavelength range. Presently, wavelength filters for daylight QKD possess a linewidth of $25.6$~pm, providing a noise suppression of over 44.7~dB across the 1380–1760~nm range, with an optical efficiency of $74.5\%$ \cite{Han2022}. To bypass this filter, we must consider the precise wavelength used by quantum communication sources, accounting for corrections due to the Doppler shift. In such a scenario, this mitigation becomes irrelevant from the perspective of a laser attack.

    \item \textbf{Spatial Filter.} The spatial filter refers to the size of the collecting telescope—a smaller telescope leads to lower noise. Despite advancements in laser and targeting technology, atmospheric turbulence remains the primary factor degrading the beam spot width, resulting in beam spreading and wandering, which can require a larger telescope or lens.

    \item \textbf{QKD Receiver Design.} In \cite{Simmons2023}, an array of single-photon avalanche diodes (SPAD) employing straightforward spatial discrimination was proposed as a counter-countermeasure effective against both in-FOV and out-of-FOV attacks. Experimental evidence supports the assertion that this solution can yield at least a 20~dB improvement in the signal-to-interference ratio compared to a~single SPD \cite{Simmons2023}.

    \item \textbf{Optical Attenuators.} Optical attenuators tailored for quantum communication incorporate a~specialised passive component—an optical fuse that permanently disconnects itself once the injected power surpasses a predefined threshold \cite{Huang2020}.

    \item \textbf{Frequency Hopping.} Frequency hopping (or agility) is a technique well-known in classical RF systems, such as radar \cite{Galati1993-mf}. The dynamic alteration of frequency (or wavelength) in optical communication, when combined with a wavelength filter, can serve as an effective counter-countermeasure against narrow-band LWS. However, compared to the RF domain, implementing this in the optical realm would present a substantial technological challenge, especially concerning photon sources and spectral filters.

    \item \textbf{Rerouting.} A natural mitigation strategy involves switching to backup Quantum Key Distribution (QKD) links or rerouting the quantum connection. This, however, can be particularly challenging in the early stages of quantum networks, where redundancy is minimal, and rerouting consumes substantial resources. For instance, a free-space QKD link employing a satellite as a trusted repeater would necessitate rerouting via a terrestrial quantum network, where repeater nodes are typically spaced approximately 100~km apart.

    \item \textbf{Alternative Methods.} Especially in the case of QKD, alternative approaches—such as ensuring a sufficient number of pre-loaded keys or adopting a post-quantum cryptographic scheme—should be available for situations where QKD is not feasible \cite{ITU-T_2020}.

    \item \textbf{Unknown Position.} Knowledge of the locations of fibre optic cables, ground optical stations, or the trajectories of quantum satellites is crucial for a successful attack. Hence, keeping such information classified and unknown is imperative.

\end{itemize}

%---------------------------------------------------------------------------
%---------------------------------------------------------------------------
%---------------------------------------------------------------------------
\section{Conclusions}\label{sec:conclusions}

This paper has focused on investigating potential countermeasures in free-space quantum communication (quantum link) with the aim of inducing a temporary or permanent quantum denial of service (DoS) from a strategic perspective, across various scenarios. The studied scenarios encompass situations where a quantum receiver, being the most sensitive and thus the most susceptible element, is located either on the ground or in space, within a satellite context. The considered attack utilises a laser situated on the ground, on an airborne platform (such as a drone, plane, or stratospheric vehicle), or in space.

The simulations conducted in this study serve primarily as illustrations, given the current unavailability of more extensive data for finer-grained simulations. Moreover, a more detailed simulation at a tactical level would necessitate the consideration of specific designs for QKD systems as well as for LWS.
Furthermore, it is worth noting that the chosen simulation mechanism tends to underestimate the threat posed by LWS, rather than the other way around. Nevertheless, these simulations provide a clear perspective on the fundamental viability of potential countermeasures in quantum communication and the resulting DoS.

Our simulations reveal that a direct in-FOV attack (where the attacking laser is within the Field of View of the target) can exert a significant impact, potentially leading to the permanent damage of single-photon detectors, even with relatively low initial power. However, the practical execution of such an attack is exceptionally challenging.

Conversely, the impact of the out-of-FOV attack is relatively modest. Nevertheless, it remains sufficient to induce a denial of service due to the elevated noise levels in the quantum channel, even with an initially very low power. In principle, the primary formidable challenge in this scenario lies in possessing an adequate targeting system for the laser, with the Ground-LEO-Ground scenario potentially presenting a notable concern.

In summary, we identify the out-of-FOV attack as the principal threat to future quantum communication. The prerequisites for a laser weapon in this context are relatively undemanding in terms of required power, and the wavelength should align with the targeted quantum communication. This implies that commercially available lasers could be employed. The foremost challenge lies in target acquisition—namely, obtaining intelligence regarding the ground station's position on one side, and effecting precise tracking and targeting of satellites on the other. This aspect stands as the pivotal parameter in discerning potential malicious actors in quantum computing countermeasures.

Based on our research, we propose the following recommendations:
\begin{itemize}

    \item Place a strong emphasis on the development of counter-countermeasures for quantum communication.

    \item Give particular priority to the advancement of receiver design, incorporating features like absorptive surfaces to minimise scattering. This should also involve the implementation of a system for detecting and localising the source of out-of-FOV attacks.

    \item As a part of the certification and validation process, it is crucial to experimentally estimate both the off-axis receiving imprint (in-scattering profile) and the scattering imprint (reflection profile) for each quantum channel receiver. This assessment helps in estimating potential risks.

    \item Maintain a policy of keeping the positions of quantum receivers (be they satellites or ground-based) as confidential as possible. This strategy enhances security.

    \item Research efforts should be directed towards exploring how laser weapons and laser attacks could be prohibited or, at the very least, limited by international agreements. This falls under the umbrella of qualitative arms control.

\end{itemize}

\section*{Acknowledgement}
I am deeply grateful to Jurgen Altmann for his invaluable expertise in laser weapons, particularly their deployment in space.

\section*{Author declaration}
Michal Krelina acknowledges the administrative support and article publishing charges provided by Czech Technical University in Prague. The author commenced work at the European Union Agency for the Space Programme (EUSPA) during the course of this research. However, it is declared that no information or data related to EUSPA has been used in this study.

\appendix

%---------------------------------------------------------------------------

%---------------------------------------------------------------------------
%---------------------------------------------------------------------------
%---------------------------------------------------------------------------
\section{Laser propagation simulation}\label{sec:appx1}

In the case of LWS, we consider the Gaussian intensity profile at a distance $z$ and at the radius $r$, given by
\begin{equation}
    I(z,r,\phi) = \frac{2P_{ini}}{\pi w_{tot}^2} \exp \left( -\frac{2r^2}{w_{tot}^2} \right) \tau_{tot} S_{tot} %\cos \phi 
\end{equation}
where $P_{ini}$ is the initial power, 
$w_{tot}^2$ represents the beam-waist radius,
$\tau_{tot}$ denotes the total optical transmittance, 
$S_{tot}$ stands for the total Strehl ratio, and 
$\phi$ is the path zenith angle.

Then, the receiver power at a distance $z$ is given by
\begin{equation}
    P_{recv} (z,r,\phi) = \tau_r\int_0^{D_r/2} dr\,r 2 \pi I(z,r,\phi) 
= \tau_r P_{ini} \left(1 - e^{-\frac{D_r^2}{2w_{tot}}} \right) \tau_{tot} S_{tot} %\cos \phi  
\quad \textrm{[W]} 
\end{equation}
where $D_r$ represents the receiver aperture (see Tab.~\ref{tab:apertures}), and $\tau_r$ denotes the optical loss of the receiver, where we assume an efficiency of one for both the transmitting and receiving parts for simplicity.

The beam waist reads
\begin{eqnarray}
    w_{tot}^2 &=& w_d^2 + w_t^2 + w_j^2
\end{eqnarray}
where $w_d^2$ accounts for the contribution from beam diffraction, $w_t^2$ from the turbulence effect, and $w_j^2$ from the jitter effect.

The effect from diffraction for a Gaussian beam is given by \cite{Gebhardt1976,Andrews_2009}
\begin{eqnarray}
    w_d^2 
        &=& \frac{M^2 z^2}{k^2 w_0^2} + w_0^2\left( 1- \frac{z}{F}\right)^2
\end{eqnarray}
where
$\lambda$ is the wavelength,
$k=2 \pi / \lambda$ is the wave number,
$M^2$ is the laser quality factor (we consider the ideal case, $M=1$),
$z$ is the distance to the target,
$F$ is focal range (we consider $F \to \infty$), 
and the initial beam waist is connected to the aperture as
\begin{equation}
    w_0 = \frac{D}{2 \sqrt{2}}.
\end{equation}

The effect of turbulence depends on whether we consider uplink (ground-to-space propagation) or downlink (space-to-ground propagation).
For downlink scenarios (i.e., $h_0>h_1$), the final turbulence waist is given by \cite{Nielsen_1994}
\begin{eqnarray}
    w_{t}^2 &=& \frac{w_d^2}{M^2} \left( \frac{D}{r_{0}} \right)^{5/3}
\end{eqnarray}
where $r_{0}$ is the Fried parameter or Fried’s coherence length, defined as 
\begin{eqnarray}
    r_{0} &=& 0.431575 k^2 \sec^{11/6}(\phi) \mu_{u/d}.
\end{eqnarray}
In the case of a downlink (i.e. $h_0>h_1$), we have
\begin{eqnarray}
    \mu_{d} &=& \int_{h_0}^{h_1} dh \, C_n^2(h) \left(\frac{h - h_0}{h_1 - h_0} \right)^{5/3}
\end{eqnarray}
and for uplink,
\begin{eqnarray}
    \mu_{u} &=& \int_{h_0}^{h_1} dh \, C_n^2(h) \left( 1 -  \frac{h - h_0}{h_1 - h_0} \right)^{5/3} 
\end{eqnarray}
where always $h_0 < h_1$.
Here, $C_n^2(h)$ represents the refractive index structure coefficient. We employ the Hufnagel-Valley model \cite{Valley1980-by}
\begin{equation}
    C_n^2(h) = 0.00594 \left( \frac{v}{27} \right) (10^{-5} h)^{10} e^{-h/1000}
            + 2.7 \times 10^{-16} e^{-h/1500}
            + A_0 e^{-h/100}
\end{equation}
where $A_0$ defines the turbulence strength at ground level, and $v$ is the RMS (root mean square) wind speed at high altitude. In the HV5/7 variant \cite{Hemmati_2021}, values $A_0=1.7 \times 10^{-14}$~m$^{-2/3}$ and $v=21$~m/s are used.

In the case of utilising adaptive optics (AO), the turbulence waist is given by \cite{Puent_2017}
\begin{eqnarray}
    w_t^2 &=& \frac{w_d^2(1-S_{ao})}{S_{ao}}
\end{eqnarray}
where
\begin{eqnarray}
    S_{ao} &=& e^{-\sigma_{ao}^2},\\
    \sigma_{ao}^2 &=& \sigma_{WFS}^2 + \sigma_{fit}^2 + \sigma_{temp}^2,\\
    \sigma_{WFS}^2 &\approx& \frac{4}{SNR^2},\\
    \sigma_{fit}^2 &\approx& \kappa \left( \frac{r_s}{r_0}\right)^{5/3},\\
    \sigma_{temp}^2 &\approx& \left( \frac{f_G}{f_{BW}} \right)^{5/3},
\end{eqnarray}
where typical values \cite{Puent_2017} are $\kappa = 0.34$, $r_s = 100$~cm, $f_{BW} = 20$~Hz, and $f_G=8-40$~Hz.
We consider high SNR, $SNR = 50$, thus $\sigma_{WFS}^2\approx0$.

The jitter effect waist is given by \cite{Gebhardt1976}
\begin{eqnarray}
    w_j^2 &=& 2 \langle \theta_{rms}^2 \rangle z^2.
\end{eqnarray}
Here, $\theta_{rms}$ defines the total root mean square angular displacement due to all contributions from the jitter. We consider $\theta_{rms} = 2.0 \times 10^{-6}$~ rad \cite{Fussman_2014}.

The total optical transmittance reads
\begin{eqnarray}
    \tau_{tot} &=& \tau_{a} \tau_t \tau_p
\end{eqnarray}
where
$\tau_{a}$ is the atmosphere transmittance computed using MODTRAN \cite{MODTRAN} for the selected wavelengths (810 and 1550~nm),
$\tau_t \approx 1$ is the optical loss of the transmitter, and
the pointing error \cite{arxiv.2208.10236}
\begin{equation}
    \tau_p = \frac{w_t^2}{w_t^2+4\sigma_p^2} \approx 1,
\end{equation}
where $\sigma_p$ is the variance of the pointing probability density that follows a Gaussian distribution.
We consider a very fine-tracking control with a very small deviation $\sigma_p$ compared to $w_t^2$.

In general, the final Strehl ratio reads
\begin{equation}
    S_{tot} = \frac{1}{1+ \sum_i( S_i^{-1} -1)}.
\end{equation}
where one could account for the thermal blooming effect \cite{Gebhardt1976}
\begin{eqnarray}
    S_{i=TB} &=& \frac{1}{1 + 0.0625 N_D^2}.
\end{eqnarray}
where $N_D$ is the thermal distortion number.
For simplicity, we consider scenarios with LWS where the thermal blooming effect can be neglected, similar to other nonlinear atmospheric effects.

The out-of-FOV attack was simulated by adapting the scattering profile model from \cite{Gozzard2021} to modify the approximated received power
\begin{equation}
    P_{recv} (z,\phi) = \tau_r  I(z,r=0,\phi) \sigma_{rec} 
  \quad \textrm{[W]} 
\end{equation}
where the in-scattering profile from \cite{Gozzard2021} reads
\begin{equation}
    \sigma_{rec} = \kappa_{outFOV}  \frac{\pi D_{r}^{2}}{4} \cos^2 \phi.
\end{equation}
Here, we introduce a parameter $\kappa_{outFOV}$, estimated as $10^{-7}$ in \cite{Gozzard2021}. For our demonstration, we consider a range of $\kappa_{outFOV}$ values, specifically $10^{-6}$ to $10^{-9}$, and a zenith angle of $\phi = 60^\circ$ (elevation $30^\circ$).

In the case of Ground-LEO-Ground, we employ the approximated optical scattering cross section (reflection profile) from \cite{Gozzard2021}
\begin{equation}
    \sigma_{sat} = S \epsilon \sqrt{\cos \phi}
\end{equation}
where $S$ is the area of the reflecting surface, and 
$\epsilon$ is the albedo. 
In our analysis, we consider a range of values for $S$, specifically $4$ and $0.1$~m$^2$, corresponding to the Micius and 1U CubeSat, respectively. The albedo is considered from maximal reflection (1) to suppressed reflection of 0.01.

It is important to note that the scattering profiles from \cite{Gozzard2021} are only very approximate and require precise estimation for each system.

%---------------------------------------------------------------------------
%---------------------------------------------------------------------------
%---------------------------------------------------------------------------

\printbibliography
\end{document}